\documentclass[namedreferences,hyperref,optionalrh]{spr-sola}
\usepackage{graphicx}
\usepackage{color}        
\usepackage{natbib}
\usepackage{url}

\newcommand{\figeven}
{
\begin{figure}[ht]
\begin{center}
\includegraphics[width=0.95\linewidth]{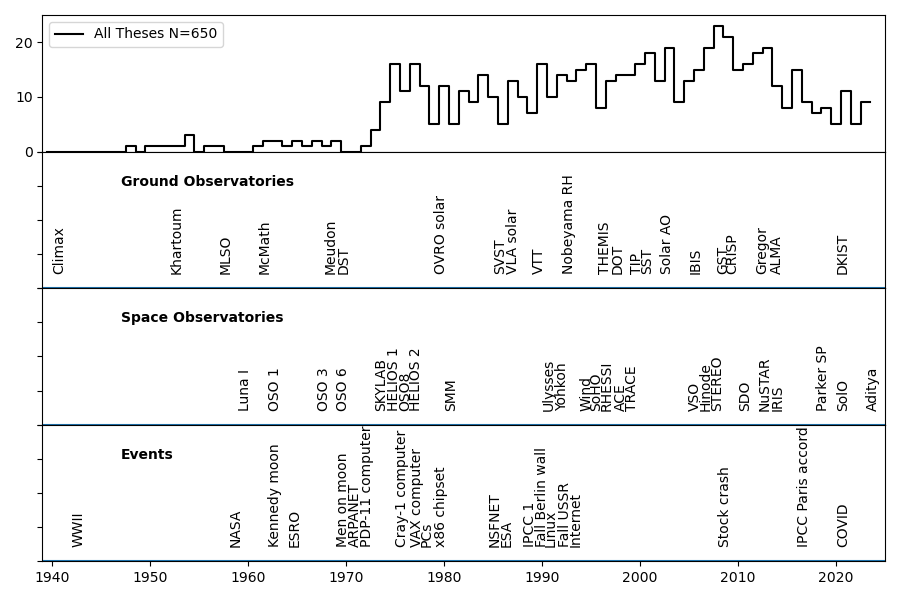}
\end{center}
\caption{The numbers of
PhD theses in our working sample are 
shown as a function of the years in which they were published.  
The lower panels indicate  
eras of various ground-based solar 
observatories, space solar 
missions, and historical events.
\textbf{Although NuSTAR is not \textit{per se} a solar mission,
it is unique in having contributed to observing the Sun 
in $\gamma$-rays and so is included.}}
\label{fig:even}
\end{figure}
}

\newcommand{\figmethods}{
\begin{figure}[ht!]
\begin{center}
\includegraphics[width=.7\linewidth]{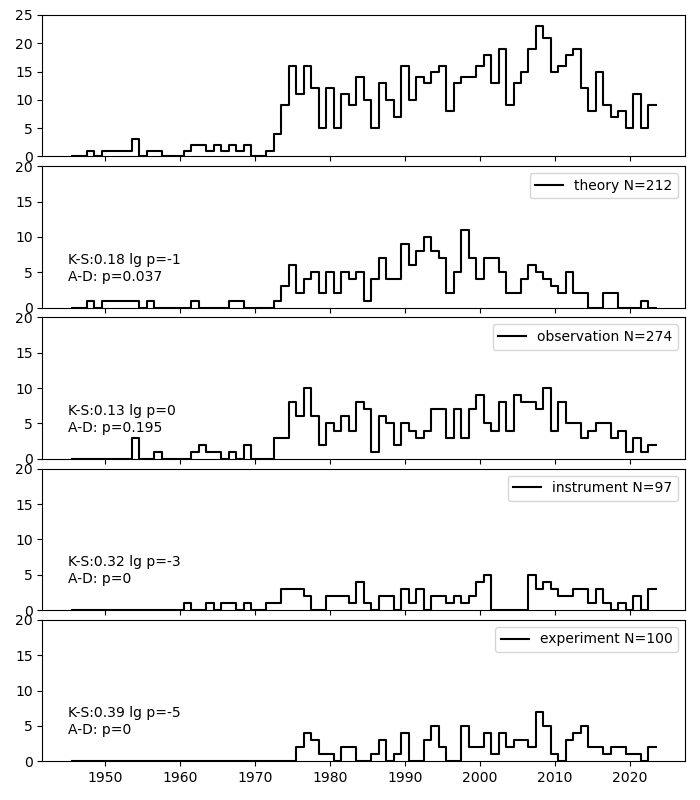}
\end{center}
\caption{The number of
PhD theses in our working sample are shown along with the methods used in the thesis work. The annotations ``K-S:''
and ``A-D:'' are followed by
rank statistical measures of
how close each distribution of the four
lower panels matches the 
sample shown at the top
(see text). }
\label{fig:methods}
\end{figure}
}

\newcommand{\figtechniques}{
\begin{figure}[ht]
\begin{center}

\includegraphics[width=.7\linewidth]{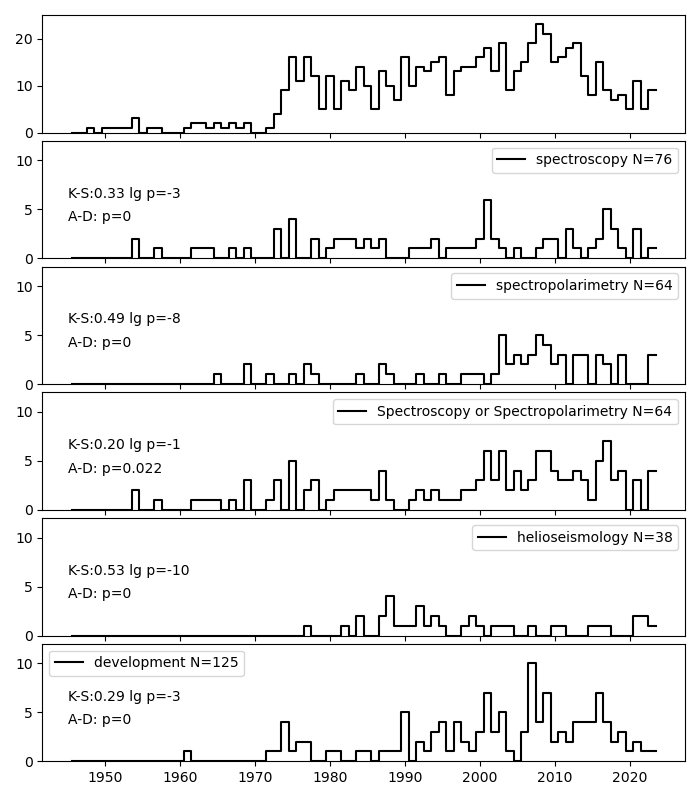}
\end{center}
\caption{Statistical distributions of theses are shown for
various techniques, as in
Figure~\ref{fig:methods}.  
}
\label{fig:techniques}
\end{figure}
}

\newcommand{\figsuccess}{
\begin{figure}[ht!]
\begin{center}

\includegraphics[width=0.95\linewidth]{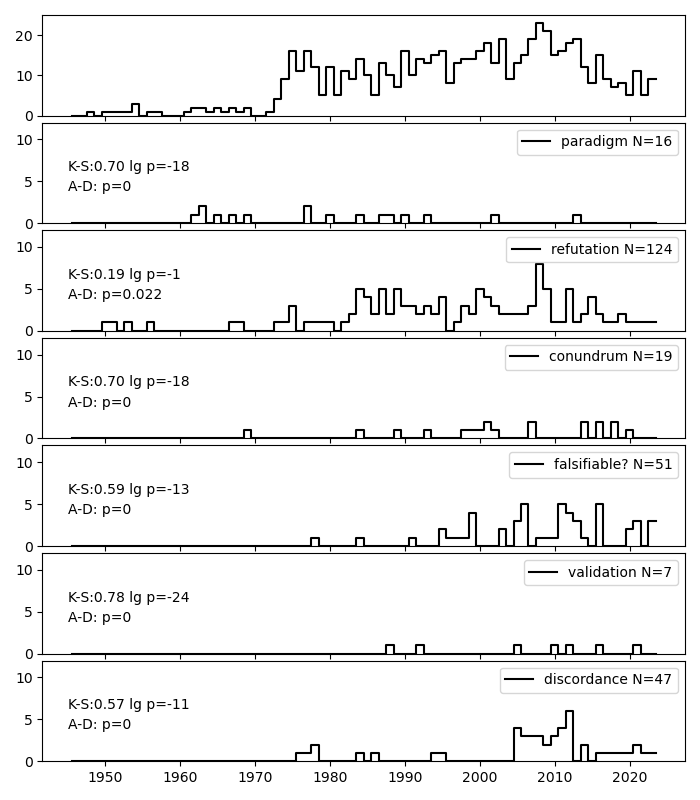}
\end{center}
\caption{The number of
PhD theses in the working sample is reproduced in the upper panel.   The lower histograms show measures of success from high to low, 
corresponding to 
keywords \textit{Paradigm, refutation, conundrum, falsifiable?, validation} and \textit{discordance.}
}
\label{fig:success}
\end{figure}
}

\newcommand{\figtargets}{
\begin{figure}[ht]
\begin{center}

\includegraphics[width=.7\linewidth]{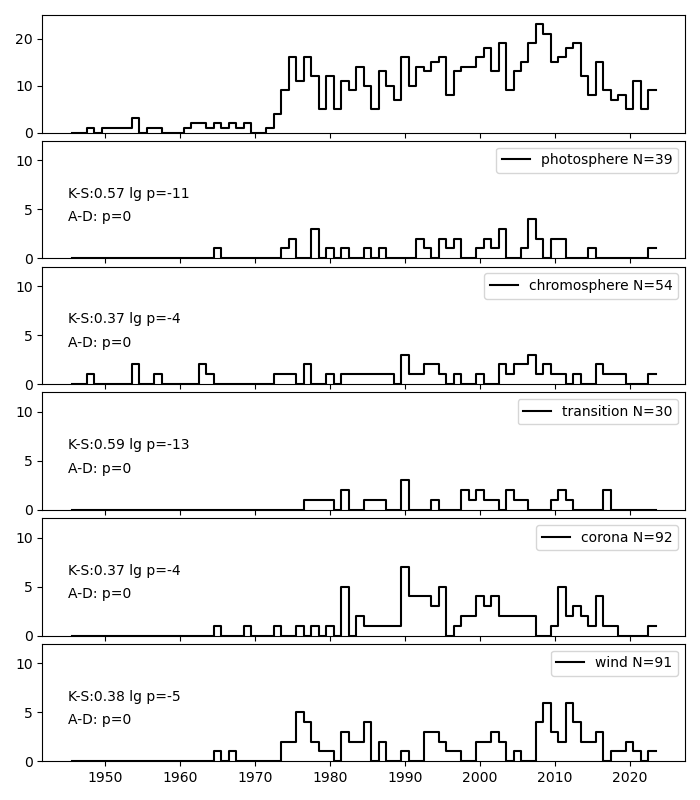}
\end{center}
\caption{The number of
PhD theses in our working sample are shown as in
Figure~\ref{fig:methods}. The A-D and K-S statistics 
show that the targeted structures (photosphere to
wind) have generally been studied more recently than 1973. Over 8 decades the distributions are probably not
drawn from the same distribution as the full sample. 
}
\label{fig:targets}
\end{figure}
}

\newcommand{\figtargetss}{
\begin{figure}[ht]
\begin{center}
\includegraphics[width=.7\linewidth]{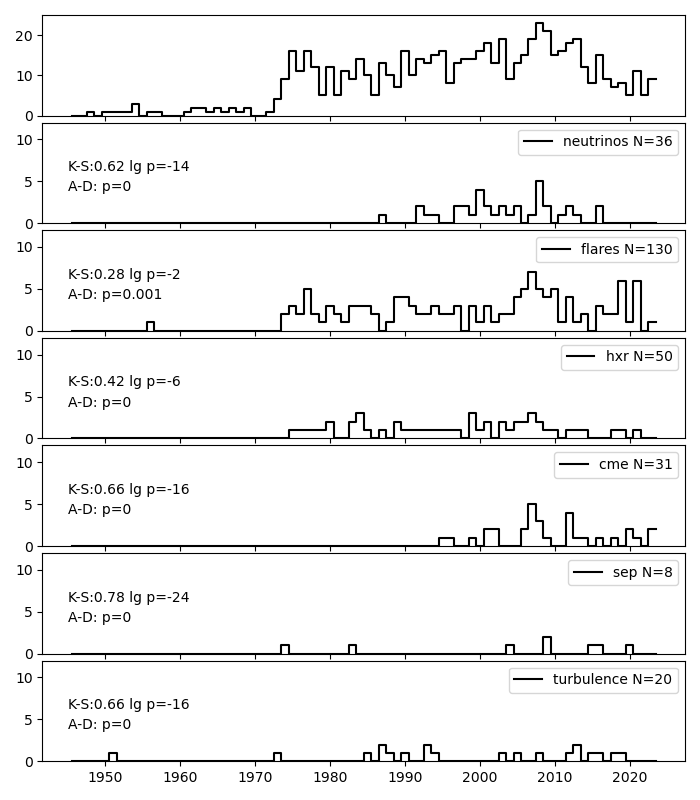}
\end{center}
\caption{The number of
PhD theses in our working sample are shown as in
Figure~\ref{fig:methods},
for more specific subjects than shown in Figure~\ref{fig:targets}.
None of the individual
subjects of study are drawn from the same distribution as the full sample.
}
\label{fig:targetss}
\end{figure}
}

\newcommand{\figalternates}{
\begin{figure}[ht]
\begin{center}
\includegraphics[width=.7\linewidth]{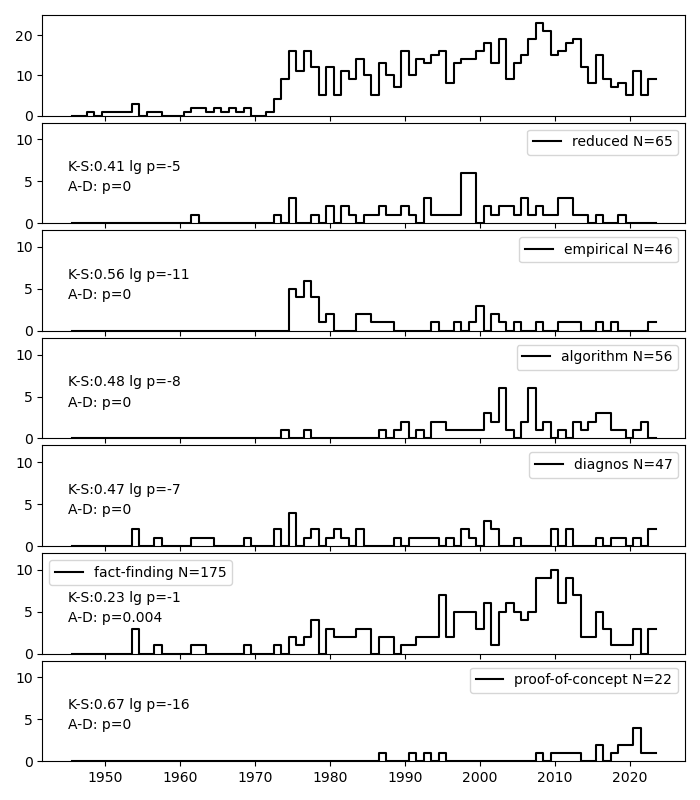}
\end{center}
\caption{Data for  alternate keyword statistics are shown.
}
\label{fig:alternates}
\end{figure}
}

\begin{document}
\sloppy

\begin{frontmatter}

\title{Changing Methodologies in Slar Physics}

\author[addressref={aff1,email=judge@ucar.edu}] {\inits{P.G.}\fnm{Philip Gordon }~\snm{Judge}\orcid{0000-0001-5174-0568}}

\address[id=aff1]{High Altitude Observatory, 
National Center for Atmospheric Research\thanks{The National Center for Atmospheric Research is sponsored by the National Science Foundation}, 
Boulder CO 80307-3000}

\runningauthor{P. G. Judge}
\runningtitle{Solar methodology}

\begin{abstract}
This study attempts to 
establish a basis for
understanding how methods used
in research in solar physics
have evolved since World War II (WWII).   The goal
is to begin to explore if and how
the changing research environment affects the training of young scientists,
and the future of solar physics research at
our institutions.
A strategy based upon a sample of 650 
PhD theses is used to seek  possible   
trends over 8 decades, 
with the aim of uncovering
any correlations between
methods used and measures of
success.  Necessarily 
subjective,  results 
depend on how methods are 
defined, and how success is measured.  Although a brief justification of the choices made is attempted, trying mainly to avoid pitfalls such as counting citations, it is clear that further assessment is required. The statistical analysis is based upon necessarily 
subjective categorization and
the inference of likelihoods of two
different 
distributions being drawn from the same
underlying distribution.  The statistics 
seem to reflect 
historical events, such as 
the Kennedy Moonshot program 
and the associated SKYLAB
mission, with changes delayed by a few years.   The data 
suggest that impactful advances 
are becoming more rare.  Yet the methods used have changed little 
barring those related to obvious technological
advances (e.g. the advent of spacecraft,
adaptive optics).   
A follow-up study to explore the 100,000+ 
 publications in
solar physics through machine learning seems
warranted.  
\end{abstract}
\end{frontmatter}

\section{Introduction}

The application of method in science remains 
a subject for debate
\citep{Newton-Smith2003}.
Philosophers may still argue over what constitutes scientific work versus
pseudo-science, but 
to practicing scientists today,  
philosophy and method are rarely of immediate 
interest.  Detailed techniques are described of course in academic articles, but the
reasons underlying methodological choices made are
discussed less frequently.   
A question then arises:
how important is it that  researchers pay attention to choosing 
methods underlying their work?
As  fresh 
post-graduates, 
many of us have little idea of what research will entail. As undergraduates, 
we have been introduced to the basics for our subject, but in deciding to pursue research, we  rightly rely on our advisors and perhaps  committees for 
guidance.    Finding a good thesis problem
is not a trivial task, yet
this choice of problem and methods to solve 
it frequently steer future career
paths.   Method is important to 
be aware of, in order that a researcher avoids pre-determined outcomes,  results of little consequence, logical fallacies, and other potential traps.   This is especially true today given the increase
in outside pressures, from excesses in publishing, 
proposal writing and programmatic 
needs, to satisfying 
a society demanding ever more direct
relevance to, and assessment of, 
a researcher's work. Consequently, 
a review of
methodologies across science 
appears worthwhile, timely and 
potentially important \citep[e.g][]{Kleppner2019}.

Here we examine methodologies underlying research
in solar physics, carried out since  
 WWII.  It is widely recognized that this war  separated eras during which  
science practitioners had radically different backgrounds, resources, and were drawn almost exclusively from only half of the population.    The ``Cold War'', which after 
WWII prompted the new era of East-West competition in science, 
fueled an unprecedented 
period of research investment.  
It set the stage for 
the scientific environment 
in, for want of  better 
description we call 
``Western Civilization'', 
bequeathed to us today. The health of a small part of Western science, solar physics, is our primary concern here.   This
obvious 
cultural bias 
is mostly a practical
one. The West
has arguably developed many more 
new tools for solar work since WWII (see Figure~\ref{fig:even},
although this plot is certainly
incomplete).  
This article concerns solar
physics in the West.

Future generations of scientists are of course 
dependent upon good mentorship.
But over recent generations, pressures upon mentors have 
evolved away from the 
na\"ive, idealistic nurturing
of investigations into Nature.  
To advance 
careers, a  
modern mentor must walk a tightrope, navigating 
teaching, mentoring, writing monthly
or quarterly reports, serving on
internal and external committees,
writing and reviewing proposals, 
articles, and managing a laboratory and/or research group. 
They live under a high pressure conditions  where ``deliverables'' become due, deadlines loom, and often arbitrary targets are to be met, set by others concerned
with institutional administration.  Consequently there is a danger 
that a well-meaning but stressed mentor 
may inadvertently be 
training their wards that research is a business. With increasing competition for resources, research may be as full of the richness and foibles of human nature as, say, car dealerships are.  
Is this ``system'' serving our students as well as it might?  
\textit{Are we training our students in 
natural science, or are we  training them to provide 
 products by specific deadlines, in response to institutional  models of management and assessment?}

 The question is of considerable importance, for there are 
many examples where fundamental research has led to 
enormous benefits to society, during eras where nothing 
of commercial or practical value was part of any measure of success
\citep{Kleppner2019}.  One obvious example
is the development of lasers, colloquially viewed at the time as ``a  solution without a problem'', another might be 
magnetic-resonance imaging. 
 
Thus we are motivated to try to identify 
common threads that underlie demonstrably successful 
research, and to find pointers useful for
budding scientists as they embark on a career of research that is increasingly burdened by external
pressures.
In an era of unprecedented and  cynical scientific skepticism, 
it is
arguably more important than ever for scientists to be seen to police themselves.  
The kind of self-assessment 
within the research community 
represented within the present study 
might serve to
 remind the public that scientists are 
their own severest critics. Advances in science  
are driven by ethical, open-minded discourse, imperfect 
as it is.  

There are many other talks I have given.  I have a comedic 

look at climate change (not political!),   I have one on some unusual 

astronomers in history (wax noses, catholic priests, rebels, 

infamous and famous people, forgotten women, astronauts),

eclipses and humanity, constellations, astrology and the age of

dquarius,... the Standard Model (of physics and the universe),

eclipses across history, Einstein and the quantum mechanics, 

what's sci fi movies got right (and wrong...)
John Archibald
Wheeler is credited with
the pithy epithet
\begin{quote}
    ``The whole purpose of science is to get the mistakes over with as quickly as possible.''
\end{quote}
Of all
human endeavors, science is the only one in which self-correction drives new knowledge and understanding. 

\figeven

\section{Methods}

Our methodology involves identification of a   representative and 
homogeneous sample of
published research from 
1945 until 2025.  A conservative estimate, based on searches for
solar physics articles in NASA's
Astrophysics Data System (ADS), 
is that over 109,000 refereed publications on solar physics 
have been published during this period.   The publications
are peer reviewed, but not homogeneously across different journals, 
or across the decades.   
Two approaches seem
feasible: on the one hand 
one might adopt machine learning
applied to the entire ADS
solar database. On the other, 
the need for homogeneity 
suggests that a more cautious 
approach is needed.  In the present work this homogeneity 
is to some degree fulfilled by
choosing to analyze 
only \textit{PhD theses}, classified ``by hand'' to try to
identify methods.   
In doing so we must  
accept the fallibility of 
human judgment. On the other hand, this seems unavoidable, because a machine must be supervised anyway to avoid 
spurious results.  A human  
has the advantage of identifying 
the essence of 
methods used by actually reading 
the literature, to 
categorize and perhaps interpret the evolving research landscape
in relation to advancing technology.  

The NASA ADS database contains 
after
screening to remove  solar-\textit{related} research (for example, solar energy cells), a 
sample of over 650  
theses on solar physics. This is
an average of nearly 
10 theses per year.   The
sample is certainly incomplete,
some theses are missing 
in cases where abstracts were unavailable in the archive (e.g.
in the 1950s). Others are  screened
out when 
abstracts contained 
work relevant to the Sun, such
as on magnetohydrodynamics, 
yet the Sun or solar phenomena
were not explicitly mentioned.
The author added 10 theses 
from the 1950s and 1960s, available at the library of
the High Altitude Observatory.

Year-by-year, such
statistics are neither interesting nor adequate.
Our interest is  
in decadal changes and longer,
time scales over which our institutions
and policies change, and 
in which paradigm-changing
results in solar physics  
seem to occur. Binned over decades, meaningful
results ought to  be possible,
with about 80 theses published per decade.  

The sample of PhD theses is also well-defined, and 
more homogeneous than 
refereed articles.  Thesis research is by definition required to be novel, meeting 
criteria that are common across many universities.
Thesis research is peer-reviewed
through external (as well as internal) examiners. The 
rules for theses vary in detail from institution to 
institution, but the overall 
requirements always include
original work by the author, 
technical proficiency (in mathematics and/or engineering, 
for example), and the derivation of some kind of
advancement in understanding,
or opening up more questions 
to advance the subject. 
In this sense,  perhaps PhD thesis work is not only 
a more homogeneous sample, but it is 
reviewed
more thoroughly 
than published articles.

\figmethods

The present study lies within  
social science:  we wish to explore
how one research area 
has evolved, both   naturally and 
in relation to  environmental pressures and opportunities.  
Social sciences face well-known 
fundamental challenges
\citep[e.g.][]{Ioannidis2005}:
sample bias, correlation vs. causation, cherry-picking and selective reporting, misinterpretation of statistical tests,
lack of reproducibility.   The 
selection of PhD theses helps
deal with bias, cherry-picking, 
selective reporting, and the 
statistical tests below are
well established and straightforward to interpret.  But this study cannot address
reproducibility, which is
why it must be considered 
preliminary as well as somewhat subjective owing to the need
for categorization.  

The set of 650 theses is necessarily incomplete, with a
strong Western bias. But completeness is not required to answer our primary questions. 
Perhaps the largest difficulty
is the modest size of 
samples per decade, particularly 
before 1973.   Below we will
also analyze results from 1970
onwards as well as 1945, 
to see if results change, with
a larger density of theses.  The
essential fact here is to acknowledge, like polling a representative sample of 100 people of a far larger population, that we must accept a random uncertainty of 10\%.   
In summary, we settled on the 
following strategy:
\begin{itemize}
    \item Search for PhD theses  that are concerned with \textit{solar physics} 
    \item Define several categories of primary  \textit{methods} adopted by the authors.
\item Perform spot-checks of these categories by reading 
thesis text or related articles, and also in consultation with several  authors.
    \item Identify external events and opportunities potentially influencing 
    research in this area.  Those 
    selected are shown in Figure~\ref{fig:even}.    
    \item Analyze the resulting distributions of thesis work according to the categories and events.
    \item Analyze the results against a different but related category set, to test for subjective bias and spurious
    results. 
\end{itemize}

\subsection{The Sample}

\label{sec:sample}

As of February 2025, our search criteria yielded 650 PhD theses about solar physics are 
listed in the ADS database. 
This represents a 
statistically significant sample, decade-to-decade,  if we accept
the 10\%{} random uncertainty.   But this noise estimate 
will not actually be used below, 
it is for context.  
Instead 
we will find more robust metrics of the degree of similarity of the 
distributions using standard methods. 

The distribution of the number of theses in the entire sample is shown 
in Figure~\ref{fig:even}, plotted 
as a function of date of publication.   Before 
analyzing this distribution and
others, first we must define 
how others are constructed.  

\subsection{Classes of Methods}
\label{sec:methods}

Each thesis was assigned 
a set of keywords to
identify methods, based upon
reading of the abstract and, 
if ambiguous, the thesis 
itself (when available). Each was assigned a number of classes sufficient to represent the overall methodology used by the thesis' author.  Each thesis may be classified 
as Observation, Theoretical, Numerical 
and Instrumental at the same time, according to a reading of the abstract.
These keywords are subjective, 
as would be the case for
a simple application of machine learning, such as clustering.
Specific examples of classifications  are given in Section \ref{subsec:ex} to reflect rationale behind the  choices made.   The reader is invited to assess these
choices for themselves. 

\figtechniques

\newcommand{\look}[3]
{\item \textit{#1.} #2. {\bf Example:} #3. }

\begin{enumerate}
\look{Observations}{The thesis was based significantly on observational
data}{``The Two State Structure of the Solar Wind and its Influence upon Proton Temperatures in the Inner Heliosphere: Observations of HELIOS 1 and HELIOS 2''}{R. E. Lopez, 1985.  
The author's thesis analyzed data from
two unique spacecraft to explore
the slow- and fast- solar wind properties \textit{in-situ} at different epochs and locations.}

\look{Theory}{The thesis studies
analytical and/or numerical 
work, perhaps including numerical experiments}{``The dynamic evolution of active-region-scale magnetic flux tubes in the turbulent solar convective envelope'', M. A. Weber, 2014}{
The thesis studies numerically the transport of buoyant
concentrations of magnetic fields
from their source beneath the convection zone to the solar surface. It is also a fact-finding exercise. The dynamics revealed in the numerical experiment is analyzed in terms of transport theory.} 

\look{Numerical-experiment}{As in a 
laboratory experiment, a numerical
experiment stands alone as a representation of a natural system whose output is to be examined for
insight}{``Modelling solar coronal magnetic field evolution'', E. E. Goldstraw, 2019}{Four different magnetohydrodynamic (MHD) approximations are computed and compared to a full numerical MHD solution. The ability of different 
approximations to represent tearing-mode instabilities in the corona is assessed both to understand 
the relative importance of physical processes and to find more efficient 
algorithms based upon simplified physics.
}
\look{Thought-experiment}{Significant analysis is devoted to drawing conclusions by deductions based on numerical or real data, without resorting to further experimentation}{``The Ionization State of the Solar Wind'', S. Owocki, 1982.  Based upon novel numerical results, this thesis argued that measurements of ion charge states at 1AU can be traced back to conditions near the 
base of the solar wind, including 
departures of electron distribution functions from Maxwell-Boltzmann statistics, identifying different behavior of 
ions with different electron shell structures}
\look{Analytical}{Primarily 
analytical theory is discussed}{``The Formation, Structure and Dissipation of Direct Electric Current Systems in the Solar Atmosphere'', G. A. Roumeliotis, 1990. 
 The author develops analytical MHD Models
describing the force and energy balance in the low corona and transition region}

\look{Synthesis {\rm or} Simulation}{In contrast to numerical experiments, in which 
solar data are not explicitly used 
in the analysis, 
a synthesis is intended to be directly compared with solar data, within 
appropriate limits}{``Numerical simulations of X rays and gamma rays from solar flares'', J. M. McTiernan 1989.  The author performs kinetic calculations 
and simulated hard X-ray spectra for comparison with solar observations, finding that some of the data cannot be reconciled with models}

\look{Conundrum}{A definite contradiction discovered previously or during thesis work formed a significant ingredient or conclusion}{ ``The Influence of Quiet Sun Magnetism on Solar Radiative Output'', C. A. Peck, 2018.  The author's work attempts to resolve discrepant data from different instruments on solar spectral irradiance variations, which differ in sign and amplitude}
\look{Fact-finding}{The author embarked upon a fact-finding mission
using novel data, techniques, including observations, laboratory experiments, and numerical experiments with a
primary goal to establish or strengthen basic
facts upon which future work might
survive scrutiny}{``Explosive Events in the Quiet Sun: Extreme Ultraviolet Imaging Spectroscopy Instrumentation and Observations'', T. Rust 2017.  The thesis analyzes 
novel data from a rocket experiment in which both spatial and spectral information is encoded in 
two-dimensional images, 
and inverted.  The data offer a unique assessment of dynamics across the three dimensions of space for the first time}
\look{Instrument-development}{
The thesis contributes significantly 
or completely to the development of
an instrument via modeling, theory, 
construction, calibration or other
activity}{``Studying Particle Acceleration in Solar Flares via Subsecond X-Ray Spikes: Analysis and Instrumentation'', T. J. Knuth,  2021.  The author
has developed instrumentation 
for rapid, precise solar X-ray instruments, a fact-finding mission exploring bursty phenomena in flares}
\look{Algorithm-development}{A substantial fraction of the thesis is devoted to developing numerical algorithms across solar physics}
{``Long-term Study of the Sun and Its Implications to Solar Dynamo Models'', B. K. Jhar, 2022}{The 
thesis describes algorithms to
extract objective measures covering data from multiple instruments over decades.}
\look{Forecasting}{A thesis with 
the goal of delivering solar behavior 
in the future}{``Solar-Flare Forecasting: a Comparative Study of Human and Computer-Based Methods'', D. M. Shaw, 1993. This is the first thesis in the database which devoted to forecasting 
solar phenomena, the author 
analyzes several approaches to
flare prediction recognizing the 
problems associated with predictions of rare events, i.e. those in the tail of distributions}
\look{Prediction}{The thesis is
devoted in part to predicting probabilities of future solar behavior
without specifically issuing forecasts to be acted upon by interested agents}{
``Innovative Methods in the Prediction and Analysis of Solar-Terrestrial Time Series'', A. J. Conway 1995.  This thesis explores the application of feed forward neural networks, and other numerical methods, to the prediction and analysis of solar terrestrial time series}
\look{Proof-of-concept}{A thesis 
that explores feasibility of new observational or theoretical 
studies with the aim of
offering fresh insight into solar
behavior}{ ``Diagnostics of solar flare energetic particles: neglected hard X-ray processes and neutron astronomy in the inner heliosphere'', P. C. V. Mallik 2010. The author explores the 
promise of 
a novel hypothesis concerning HXR emission during flares, as well as gamma-rays
and neutron detection, by building models to predict fluxes detectable from space}
\look{Calibration}{A significant fraction of the thesis was devoted to the calibration of instruments 
which measure solar phenomena, via remote sensing or \textit{in-situ}}{``Absolute velocity measurements in the solar transition region and corona from observations of ultraviolet emission line profiles'', D. M. Hassler, 1990.  The thesis discusses the \textit{absolute}
wavelength calibration of a UV rocket spectrograph to an unprecedented $\pm 1$ km~s$^{-1}$, which are far smaller than systematic
differences measured  between
different spectral lines}

\look{Correlation}{The thesis 
was concerned with statistical correlations 
from real or numerical experiments, a large part of a fact-finding exercise devoted to
laying down foundations against which theories and models can be tested}{``An observational analysis of the solar chromospheric continuum intensity'', K. N. Karlsen, 2003.  
Line-of-sight magnetic fields are spatially correlated with 
UV continuum brightness, as a fact-finding endeavor. The 
study took advantage of
an unprecedented 
quantity and quality of
co-aligned data from
diverse instruments}
\look{Development}{The thesis 
is in part devoted to developing novel instruments, hypotheses or 
algorithms}{``Study of solar Ultra-Violet and X-radiation'', K. P. Pounds, 1961.   The author
describes three novel instruments for measuring radiation, prior to launch}

\look{Machine-learning}{Novel application and/or development of machine-learning techniques are 
central to the thesis}{``Automatic solar feature detection using image processing and pattern recognition techniques'', M. Qu, 2006.  Machine learning tools (Multi-Layer Perceptron,  Radial Basis Functions, and Support Vector Machines) were used to classify and correlate features in solar data, prior to studying dynamical evolution}
\end{enumerate}

\subsection{Measures of Impact}
\label{sec:success}

In a subjective order of decreasing impact, keywords measuring impact are:
\begin{enumerate}
        \look{Paradigm-shifting} {The work had a demonstrable effect on future research}{
    ``A New Picture for the Internal Rotation of the Sun", by C. A. Morrow, 1988.  The thesis was among the first works  to 
    determine the global rotation profile of the solar interior}
    \look{Refutation}{The author clearly rejects specific hypotheses or models, whether by real or numerical or thought experiments}{``X-Ray Polarimetry: the Measurement of the Polarization of Solar Flare X-Rays and the Design of a Compton Polarimeter'', by L. J. Tramiel, 1984}{Deployment and calibration of a novel instrument detected 
    no polarization in the 5-100 keV
    X-ray range from multiple flares,
    with implications for subsequent models.}
    
    \look{Falsifiability?}{This keyword flags research which 
    is not clearly falsifiable}{
    Representative examples include extrapolations,
    comparison of data with 
    models far from solar regimes, the application of
    force-free fields to
    non force-free magnetic fields, under-constrained inversions of information-poor data}

\look{Validation}{Thesis work aims to confirm previous 
ideas, interpretations or theories}{``A solar extreme ultraviolet flux model'', K. Tobiska, 1988.   A model of
thermospheric variations based upon two timeseries measurements of UV solar irradiation is validated against thermospheric density measurements through the decay of orbits of artificial satellites}

\look{Discordance}{This keyword  
is attached to studies in which there is no clear
separation between a hypothesis/theory, and experimental data}{A hypothetical example 
might be the use of data assimilation to update 
the state of a model calculation.
Another might be to use a static model to
``invert''  solar spectral data, knowing also that information must be added
to deal with non-uniqueness.  In either case refutation is difficult because of a hierarchy 
of implicit, untestable  assumptions}

\end{enumerate}

The reader may wonder why the number of citations is not   listed.    PhD theses are generally not cited, but  
associated papers are.   Other 
serious problems with counting citations 
are well known \citep{Worrall+Cohn2023}, and
should merit careful attention
in any follow-up study.

\figsuccess

\subsection{Subjects}
\label{sec:subjects}

Keywords were
assigned to the main subjects within 
each thesis:  \textit{Balloon, Bright points, Chromosphere, Convection, Corona, Coronal Mass Ejections (CMEs),  Dynamo, EUV, Flares, Fundamental physics, Gamma rays, Hard X-rays (HXR), Helioseismology, Infrared, Instrumentation,
In-situ measurements, Kinetics, Machine-learning, Magneto-hydrodynamics (MHD), Neutrinos, Oscillations, Photosphere, Polarization, Prediction, Proxies, Radio, Reconnection, Rocket, 
Solar Cycle,  Solar Interior, Solar Wind, Space Weather, Spectral diagnosis, Spectropolarimetry, Spectroscopy, Solar stellar, Stability, Transition-region, X-rays.}  While
not related to methodology \textit{per se}, these keywords 
serve to seek if differences 
exist between sub-fields of 
solar study. Representative examples  of such keyword distributions are shown in an
Appendix.

\subsection{Examples of Subjective Keyword Assignment}
\label{subsec:ex} 

The kinds of arguments used 
for specific keyword assignment 
are summarized here for two
theses out of the 650.   The reader can again assess the credibility  of the subjective 
methods adopted. 

The following thesis 
earned multiple keywords on the basis of
its abstract: ``Waves in Inhomogeneous Moving Media with Application to the Solar Wind'',  S. A. Jacques, 1977.
The abstract reads 
\textit{A complete set of equations are derived for an interacting MHD system composed of waves and a slowly varying background fluid through which they propagate. The equations for the background state are valid for any MHD wave model and conserve both the total momentum and the total energy of the system. Also derived are the dispersion relation and the equation of conservation of wave action, which governs the wave amplitude. The theory is then applied to the solar wind, with particular attention to high-speed streams. Several models are constructed for both the high- and low-speed solar wind for various values of the Alfven wave energy flux at the inner boundary (1 solar radius).}

The keywords assigned to this thesis were 
\textit{Theory, Analytical,  Numerical experiment, Thought experiment, MHD, Solar wind,  Paradigm changing}.   Arguably the only ``subjective''
assessment,  paradigm changing. 
A weak justification is that the 
thesis presents a formalism based upon a mildy 
inhomogeneous background state in which MHD waves propagate.  The thesis presents the first equations to express
conserved quantities in their regime of
applicability.   Later workers 
adopted the results for decades.  In stellar research it remains a standard framework in which 
winds of stars with surface convection are discussed.  

Another example is provided by
the thesis 
 ``Are Solar Emerging Flux Regions Carrying Electric Current?'', 
 K. D. Leka,  1995.  The abstract reads \textit{
 Flare-productive active regions exhibit non-potential magnetic field structures, 
 often described as `sheared' or `twisted' fields. This 
 morphology indicates that electric currents are present. 
 In this thesis I test whether surface flows generate observed active-region currents, or whether these currents are produced prior to their appearance at the surface as sunspots, i.e., deep in the solar convection zone. 
 To study this question I observed emerging magnetic flux in a uniquely rapidly growing active region. First I undertook an exhaustive study of the more than 50 bipoles which appeared in a sunspot group visible in August 1992. 
 I determined the time of emergence, magnetic connectivity and patterns of overall development of this young active region. Then, four independent analysis methods were used to determine whether the emerging flux was carrying the electric current prior to its appearance, or if the observed strong currents were generated by plasma flows in the photosphere.
 The four approaches gave consistent results. For a few young bipoles, I show that the morphology of chromospheric and coronal loops were definitively non-potential, that those same dipoles had proper motions which reflected twisted subsurface flux bundles, 
 that electric current existed in greater abundance than could be generated given the observed characteristics and finally that the electric current increased as the magnetic flux itself increased with no substantial delay. 
 All evidence was also consistent with a direction of twist defined by $J_{z}/B_{z} < 0$. This twist direction was also present in the older flux of this active region. I conclude that the electric currents observed in this solar active region were not produced by plasma motions in the photosphere. 
 Rather, the evidence presented in this thesis supports the hypothesis that active region electric currents are generated either deep in the convection zone or are produced with solar magnetic fields in a dynamo process.}   
 
The keywords assigned were
\textit{Observations, Spectropolarimetry, Magnetism, Photosphere, Refutation, Polarization, Fact-finding}.   The author kindly confirmed via private communication (2025) that these keywords were not inappropriate.

\subsection{Statistical Methods}

Having taken steps to
avoid some common pitfalls
in sampling in social
science (Section \ref{sec:sample}), the core method
used is to compare
two different distributions as the numbers evolve in time. 
Two non-parametric statistical tests are
applied: the classical
Kolmogorov-Smirnoff 
test, and the 
Anderson-Darling test 
 \cite{Anderson+Darling1952}. In each case, a
measure of   distance between the cumulative  distribution functions of a K-S test and the  full sample distribution is computed. For both the K-S and A-D tests a probability measure that the two were drawn from the same distribution is given (log${10}$ $p$ and $p$ respectively), these numbers are shown near $x=1950$.  As rules of thumb,
two distributions can 
be reasonably deemed to be drawn from the \textit{same} distribution 
when the K-S probability exceeds
0.05, and the A-D probability exceeds 0.001.    

\section{Results}

Figure~\ref{fig:even} shows 
the number of theses published
per year between 1945 and 2025 
(January).   Underneath the histogram,
the advents of new facilities are indicated.  The lowest panel
indicates some  potentially relevant events which may potentially  
influence
research in solar physics. 

The dominant feature of
Figure~\ref{fig:even} 
is the rapid growth and sustained jump beginning in 1973. The growth was surely prompted by  the accumulated developments 
in the cold war (the ``space  race'') during the 1950s and 1960s, Western culture being 
hugely influenced by NASA's 
manned spaceflight program culminating in the 
Apollo missions, including the 
SKYLAB mission.  SKYLAB carried 
a spectacular set of solar observing instruments 
operated by astronauts.  
This era witnessed enormous growth
in development of 
 space and ground-based 
observatories and instruments, in continued 
advances in computing hardware, 
optics, in magnetohydrodynamics, plasma
physics, theoretical spectroscopy, and radiative transfer.   The
apparent drop in number of theses 
after 2014 may in fact reflect
a maximum near 2012- the data
are too sparse to draw conclusions.

\subsection{Statistics of Methods and Techniques}

Figure~\ref{fig:methods} shows
statistics for the ``methods''
keywords (section~\ref{sec:methods}).
The data suggest that both 
theory and observation 
are drawn from the same distribution 
as the full sample, according to 
the computed probabilities.
In contrast, instrumentation and experimentation seems to be 
a significantly later development.  The statistics 
evaluated using only data from 1970 onwards are not statistically 
different from the overall
sample over the same range of dates.   We tentatively  
conclude
then that there has been \textit{essentially no change
in the methods used in solar physics research over the last 5 decades}.

Figure~\ref{fig:techniques}
shows statistics for
 keywords describing 
specific techniques in
solar physics.   Formally, all  individual distributions 
are statistically different from
the full sample,  with the possible exception of ``development''.   This keyword
encompasses work in which algorithm, theoretical, analytical, instrumental or experimental development 
was intrinsic to the thesis.  
By combining spectroscopy and 
spectropolarimetry, a distribution is found that is 
compatible with that of the full sample, suggesting an 
evolution in research techniques 
from intensity spectroscopy
to Stokes polarimetry, from
early work in the 1960s.   The
motivation, of course, was
the need to understand 
the Sun's evolving magnetic field, a goal unchanged from 
W. O. Roberts' mission to Climax 
observatory in 1940. 

\subsection{Statistics of Measures of Success}

Figure~\ref{fig:success} 
contains the data of most 
importance to this study.  Only   ``refutation'' seems to be consistent with the whole distribution, with 1/4 or the theses clearly refuting 
an hypothesis, model or
theory.    The ``paradigm'' shift 
distribution is weighted towards 
the 1960-1980 era, with just four theses deemed as 
paradigm changes since 1990:
``Soft x-ray/extreme ultraviolet images of the solar atmosphere with normal incidence multilayer optics'', 1990, L. F. Lindblom
(first successful application
of multi-layered optics to 
EUV solar imaging); ``Dynamic Evolution of Emerging Magnetic Flux Tubes in the Solar Convective Envelope'', 1993, Y. Fan (the dynamics of unstable magnetic flux ropes generated deep in the Sun is revealed
in a numerical experiment); ``Model-independent measurement of the neutral-current interaction rate of solar boron-8 neutrinos with deuterium in the Sudbury Neutrino Observatory'', K. M. Heeger, 2002 (the long-standing ``solar neutrino problem'' is show experimentally to be 
a problem in elementary physics'');
``Combining Models of Coronal Mass Ejections and Solar Dynamos'', J. Warneke, 2013 (a 
coupled interior-corona magnetic model reveals surprising physical 
connections across the entire
Sun). 

Those classified as ``conundrum'', ``falsifiable?''  ``validation'' and ``discordance''  
tended to occur mostly after 1990.   Conundrums indicate work in which a definite contradiction prompted research or emerged from the research. This is a highly desirable quality, almost always leading to a revised understanding or perhaps even to a paradigm change 
\citep{Kuhn1970}.  

The lower three rows of Figure~\ref{fig:methods}
potentially show a post-1975 evolution  towards some  
methodological problems.  
The statistics suggest that almost 15\%{} of theses published since 1975 are based upon ``discordant'' methodologies, those
without a clear separation between
experimental data and conjectures, hypotheses, theories or models.   In principle, theses in  this
sub-sample will be 
difficult to falsify.   Theses seeking primarily to validate
methods, software, hardware or
calibrations for example, 
are not necessarily strengths
or weaknesses in methodology,
instead this might be considered
as tool development, an important activity in research.   However,
if a thesis fails to refute a
conjecture or model, for whatever
reason, then the work may have 
less impact.

Of these three keywords, the 
``falsifiable?'' classification is arguably the most subjective.  
The problem is that some skill is needed to understand the outcome of
such efforts, which runs the risk of increased subjectivity, because there is no clear challenge of hypotheses or models by data.  Both are often entwined, convoluted.  In reading 
representative samples of
such work, one finds perfectly reasonable justifications for the methods used, because sometimes  such 
mixing might be required to make any progress at all.
An example might be the adoption of an 
electron beam (highly collimated tail of the electron distribution function) from the outset of an investigation of hard X rays in flares. Yet this assumption, which may be correct, has never been 
tested directly using information-rich data, such as polarization of light resulting from such an extreme anisotropy.  Other scenarios have not been refuted \citep[e.g.][]{Fletcher+Hudson2008}.
But herein lies an opportunity. Perhaps there are measurements to be made which more directly test such hypotheses, and instruments to be built.  

The role of falsifiability 
remains controversial 
even though it was formulated almost a century ago \citep{Popper1934}. Related 
is the use of Bayesian statistics where posterior probabilities replace the binary choice of true or false as an outcome.  A prior
probability must be assessed 
against the posterior probability,
guaranteeing a degree of separation between hypothesis (via the prior) and data (yielding
the posterior).   
Of the solar theses just three explicitly called out Bayesian methods in 2011, 
2015 and 2016\footnote{The editor kindly pointed to a thesis not included in the ADS
database from 2019: 
``Bayesian Analysis of the Solar Corona'', M. Montes-Solís, 
\url{
https://cloud.iac.es/index.php/s/ZwFGXK4xtK7k2GC}
}

This is
perhaps something worth
considering in the future. 
The argument against such rigor
was made famously by 
 \cite{Feyerabend1975}.
The issue of falsifiability 
is of course not resolved here,
it is meant only to flag  possible  methodological issues.

\subsection{Statistics of Alternate Classifiers}

\figalternates
Finally, Figure~\ref{fig:alternates}
shows alternative methodological keywords
to those forming the primary analysis.   These were independently assigned by the author, they have slightly different meanings to those discussed already.
``Empirical'' is a keyword expressing 
a mix of assumptions and 
data, for example a semi-empirical atmospheric model
based upon smooth temperature distributions with height, in hydrostatic equilibrium, was commonly 
developed before the 1990s to deal with radiation transfer in strong spectral lines. These methods are responsible for the increase occurrence rate in the period 1975-1981, where 
the parameters were adjusted to fit certain 
critical observations. Another example is the use of
line ratios with a host of
implicit assumptions to 
estimate average plasma 
properties such as temperature, density, or  elemental abundance
in the Sun's atmosphere.   The 
``empirical'' classification is  therefore similar to 
keywords for discordance and falsifiability.  
A related keyword ``reduced'' indicates that the analysis 
depends on reduction of the number of degrees of freedom, for example, by 
using reduced MHD or one dimensional dynamics.   The associated problems can be 
to remove entire classes of solutions 
to dynamical equations 
associated with an ignorable coordinate. One example
is called out by \cite{Parker1991phase}.

Proof-of-concept 
is an important contribution in that 
work looks to the future to 
make advancement.  
``Fact-finding'' is the only alternate keyword for which the alternate statistics seem to be 
drawn from 
the full sample's distribution.   Although not truly independent (because the keywords are subjective in the mind of the author), there is at least some internal consistency demonstrated by the statistics of Figure~\ref{fig:alternates}
and earlier figures. 

For reference, an appendix shows data for 
specific solar targets (e.g. photosphere) and for other
phenomena (e.g. flares).

\section{Conclusions}

Some preliminary indications of changing methodology in solar physics have been derived by comparing the statistics of keywords categorising 650
PhD theses since 1945. The statistics reveal both 
unchanging and changing methods adopted by the 
authors over decades, some clearly depending on technology development 
(e.g. neutrino detectors),
others evolving without 
clear reasons.  The 
data show that breakthroughs which change the future direction of research are becoming rarer. However this may reflect the natural increasing maturity
of our understanding of a long-studied natural system,
as much as any obvious 
weakness.   On the other hand,
this work might suggest the need for a sharper focus on the many
remaining fundamental problems, 
with an eye to obtaining 
ever more information-rich data to
challenge current understanding.  
There remain many new opportunities.

There are 
potential concerns: statistics 
suggest 
increasing occurrences of
discordant methods, intrinsically mixing 
observations and theory, possibly facing 
logical fallacies.  Similarly, a modest rise in studies with 
un-falsfiable outcomes is 
concerning, perhaps 
leading to results of little consequence.  But these are
just warnings, not meant as
criticisms.   

Re-emphasizing, this  \textit{preliminary study} is intended to prompt a larger 
survey, most probably using machine learning techniques. 
Some difficulties in extending
this study to over 109,000 
publications in the ADS database
were discussed.   The inhomogeneous
nature of refereeing across 
a multitude of journals is of concern, as
well as the identification 
of methodological strengths and weaknesses using automated techniques, based solely upon 
paper titles and abstracts alone.
Nevertheless it is hoped that
others will recognize the potential
importance of studying 
methodology in the face of
modern challenges, both as 
formal students, advisors, mentors
and management, even though we
are all fortunate to be students of the Sun.   

\newcommand\apj{Ap. J.}
\newcommand\solphys{Sol. Phys.}
\newcommand\ao{Appl. Optics}

\begin{fundinginformation}
The author is grateful for support through the National Science Foundation. 
This material is based upon work supported by the NSF National Center for Atmospheric Research, which is a major facility sponsored by the U.S. National Science Foundation under Cooperative Agreement No. 1852977.
\end{fundinginformation}

\section*{Appendix A: Selection of Solar Physics Theses from NASA's ADS}

\textbf{The first 
and most important step
in this analysis is the selection of the 
sample.  A second but important 
criterion is to find such a sample containing sufficient
data to perform classifications.
After an initial search for online data we adopted  NASA's ADS resources.
}
We used the ADS to return 
all PhD theses using the following syntax in the search  string:

\begin{verbatim}
        doctype:phdthesis and title:(solar OR sun) 
        -title:( word1 or word2 or ...)
\end{verbatim}

The words or phrases 
that, if present in the title
of the publication, were used to
reject the publication from our working list, are as follows.   The searches are case-insensitive. 
The number of perceived breakthroughs 
preceding this growth
can be seen in the second 
row of Figure~\ref{fig:success}, listed under ``paradigm'', meaning paradigm-setting.
The text in this and following figures, such as ``A-D:0.64",  indicates a non-parametric ``average distance'' between the paradigm-setting and full sample  distributions shown.  This is measured using 
a metric developed by 
\citet{Anderson+Darling1952} which finds 
average normalized distances between the 
two cumulative distributions.  A number 
close to zero indicates a statistical similarity
between two distributions, above about 0.2 they 
are different.   
The better-known Kolmogorov-Smirnov statistics -- in this case ``K-S:0.71 lg p=-19" --  see 
\cite{Chakravati+Laha+Roy1968} are 
the distance parameter (0.71) and 
probability ($10^{-19}$) that the two distributions  
have been drawn from
the same parent distribution (the null hypothesis).  
The K-S test returns the largest normalized distance 
between two cumulative distributions. 
The two-sample version of the K-S test was applied here. 
Usually a value larger than $p=0.05$ indicates no significant difference between distributions.
These statistics are repeated for all panels in the Figure,  which represent different degrees 
of scientific success, 
in order of decreasing success. 
The lowest two
rows are more subjective in nature, they are considered to 
be less successful than the 
upper two rows.  Validation is more of an exercise in engineering or
technology, and if the results are questionably falsifiable,
then there must remain doubt
concerning their value.

One of the statistically significant results is the small
 difference between the full sample and the distribution of theses flagged
by ``refutation'', with $p=0.1$.  In words,  refutation is
a measure of success which is drawn from the 
sample distribution without changes over 8 decades. 

\textbf{Here is the list of 
phrases, consisting either of more than one word enclosed in quotation marks, or of individual words, which frequently accompanied 
the word ``solar'' but which, upon examination, were clearly 
not concerned with solar physics
\textit{per se}:}
\textit{
``extra solar"
``f region"
``gauge theories"
``heat transfer"
``interstellar medium"
``population i"
``solar energy"
``solar power"
``solar radiation"
``solar system"
``star forming"
``su("
``thermal energy"
``thermal systems"
advisor
aerosol
aerosols
agriculture
airglow
anaerobic
anthropogenic
asteroid
aurora
automotive
autonomous
banks
binaries
biological
biomass
buildings
cells
cerenkov
chimney
chlorophyll
chrome
cirque
city
climate
cloud
cloudy
clusters
collector
communities
composite
concentrating
consumption
control
conversion
culture
desalination
diesel
diurnal
dwellings
early 
ecology
economies
electrification
engines
entanglement
exchanger
exospheric
farming
foci
fuel
galaxy
generation
geomagnetic
geospace
giant
goals
grains
gravity
greenhouse
grid
heaters
helium-burning
homes
inverter
ionosphere
ionospheric
jovian
logic
magnetosheath
magnetosphere
magnetotail
mars
meteor
methane
middle
mitigation
moon
nanomaterial
nanoparticles
nebula
neighborhood
neptune
oregon
organic
oxide
ozone
panels
passive
pelotint
photocatalyst
photochemistry
photolysis
photovoltaic
physiology
pipe
planet
plants
political
ponds
pump
pumped
pv
receivers
renewable
residential
ribless
sea
semiconductor
social
solar-blind
still
stratosphere
supergiant
supply
thermoelectric
thermosphere
tides
tropospheric
turbines 
utility
vehicles
water
weather
winter
zenith
zonal}

Of these rejected keywords,
of note is ``weather'' -- as of February 2025 the ADS contains almost 42 theses with titles including ``space weather'' (but excluding ``weathering''),  beginning in 
1997 (``Space weather physics, prediction and classification of solar wind structures and geomagnetic activity using artificial neural networks''
Wintoft, Peter, Lund University, Sweden).

In short, the
sample cannot be described 
as ``representative'', but it is hoped that the results are ``indicative'' of 
the development of research in 
solar physics.

\section*{Appendix B: Statistics of Targets}

Statistics of some of the most common targets of PhD solar research, those for which the plasmas are directly observable,  are shown 
in Figure~\ref{fig:targets}.
While not directly related to
methodology, this plot 
is interesting from several
points of view.  Solar flares were the subject of over 20\%{} of all solar work.   Curiously, each of the corona and
wind occupied 14\%{} of PhD 
studies, whereas  the
transition region, chromosphere and photosphere
together amounted to 19\%.   The reader can judge for herself the significance or otherwise of such plots.  The sum of all the 
targets observed is $N=306$, a little less than half of all
solar theses.   Figure~\ref{fig:targetss}
shows a similar plot for 
various solar phenomena.

\figtargets

\figtargetss

\bibliographystyle{plainnat}
\bibliography{refs} 

\end{document}